\documentclass[prl,twocolumn,superscriptaddress]{revtex4-1}
\usepackage[T1]{fontenc}
\usepackage[latin9]{inputenc}
\usepackage{color}
\usepackage{amsmath}
\usepackage{hyperref}
\usepackage{amssymb}
\usepackage{graphicx}
\usepackage[english]{babel} %
\usepackage{physics} 
\PassOptionsToPackage{normalem}{ulem}
\usepackage{ulem}

\makeatletter
\@ifundefined{textcolor}{}
{%
 \definecolor{BLACK}{gray}{0}
 \definecolor{WHITE}{gray}{1}
 \definecolor{RED}{rgb}{1,0,0}
 \definecolor{GREEN}{rgb}{0,1,0}
 \definecolor{BLUE}{rgb}{0,0,1}
 \definecolor{CYAN}{cmyk}{1,0,0,0}
 \definecolor{MAGENTA}{cmyk}{0,1,0,0}
 \definecolor{YELLOW}{cmyk}{0,0,1,0}
}

\usepackage{braket}
\usepackage{bbold}

\makeatother

\usepackage{babel}
\begin{document}

\title{Imitating non-equilibrium steady states using time-varying equilibrium force \\ in many-body diffusive systems}

\author{Ohad Shpielberg}
\email{ohad.shpilberg@college-de-france.fr}
\email{ohad19@gmail.com}
\affiliation{Coll\`{e}ge de France, 11 place Marcelin Berthelot, 75231 Paris Cedex 05 - France.}

\author{Takahiro Nemoto }
\affiliation{Philippe Meyer Institute for Theoretical Physics, Physics Department, \'Ecole Normale Sup\'erieure \& PSL Research University, 24 rue Lhomond, 75231 Paris Cedex 05, France}

\selectlanguage{english}%

\begin{abstract}

An equivalence between non equilibrium steady states (NESS) driven by a time-independent force and stochastic pumps (SP) stirred by a time-varying conservative force is studied for general many-body diffusive systems. When the particle density and current of NESS are imitated by SP time-averaged counterparts, we prove that the entropy production rate in NESS is always greater than that of SP, provided that the conductivity of the particle current is concave as a function of the particle density. Searching for a SP protocol that saturates the entropy production bound reveals an unexpected connection with traffic waves, where a high density region propagates against the direction of the particle current.

\end{abstract}

\maketitle

\section{Introduction}

Long lasting non-equilibrium behaviour can be the result of many different physical scenarios, ranging from quenches \cite{Calabrese2016,Derrida2009,Smith2018}, to rugged energy landscape  \cite{Ritort_Sollich,Berthier_Biroli,aarts1988simulated} and even active systems \citep{Marchetti2013review,Prost2015review,Cates2015Motilityinduced}.  
Among these, presumably the simplest non-equilibrium  class   
is the non-equilibrium steady state (NESS). A NESS can be created by subjecting a physical system to a time-independent driving force, such as a chemical and electric potential, pressure or temperature gradient. Theoretical understanding of simple NESS systems, strongly motivated by the application to biological systems,  has greatly advanced in the last decade shedding light on the nature of fluctuations \citep{Barato2015,Cates2015Motilityinduced}, non-equilibrium phase transitions \cite{Vicsek1995,Chate2004,PhysRevE.74.061908} and even large deviations \citep{Touchette2009LDFreview, Bertini2015,Derrida2007}.

A different class of non-equilibrium behavior is observed when the system is subjected to a periodically time-varying force. These forces are often assumed to be the gradient of a potential, which ensures that while the system is out of equilibrium, freezing out the potential at any arbitrary time allows the system to relax to equilibrium \citep{Rahav2008,Chernyak2008,Maes2010,Asban2014,Rahav2017}. In the context of molecular motors, the study of time-periodic forcing has been inspired by experiments, aiming at creating  artificial molecular machines  \cite{Leigh2003,ErbasCakmak2015,Coskun2012}, known as stochastic pumps (SPs) \citep{Astumian2018}. 
While many molecular motors in nature prefer having NESS dynamics, SPs have an advantage over NESS when designing artificial molecular machines~\footnote{For example, varying the temperature in time is much more realistic than coupling a micro-size system to two different thermal baths.}. Towards the goal of building efficient molecular machines, it is important to study what are the limitations of SPs when designed to imitate a NESS machine.

Driven by this question, Raz {\it et al.}  studied an equivalence class between NESS and SPs in a master equation on a {\it discrete-space} \cite{Raz2016}. 
They proved that the fundamental properties of the NESS system, defined as the steady state current, probability distribution and entropy production rate \footnote{The triplet probability, current and entropy production rate completely defines the master matrix \citep{Raz2016}. In that sense, they are fundamental. }, can be exactly mimicked by the corresponding  time-averaged quantities of a SP protocol.  
A similar question has been addressed in a {\it continuous}-space system limited to the dynamics of   a single Brownian particle subjected to an external force confined on a ring. 
However, contrary to the master equation description, a complete mimicking has failed. Instead, they found 
an inequality between the entropy production rates of NESS and SPs when the rest of the fundamental properties are exactly mimicked \citep{Busiello2018}: SPs entropy production rate is always greater than the NESS counterpart. 
Furthermore, they also found that this inequality can be asymptotically saturated by using a protocol that entails approximating a delta-function traveling wave.  
The bound and the protocol work for any forcing of the Brownian particle, but the dynamics beyond the single particle picture has not yet been addressed.

In this paper, we consider a class of {\it many-body} diffusive dynamics confined on a ring, and we extend the inequality and the saturating protocol obtained in \cite{Busiello2018} to these systems. To this goal, we first show that for diffusive systems,  the entropy production bound exists as long as the (density-dependent) conductivity is concave. We then argue that the universal protocol found in \cite{Busiello2018}, is generally limited to non-interacting particles (or the single particle picture). To overcome this difficulty, a different scheme is devised to   saturate the bound for concave conductivity models in the case of  a constant NESS force. Surprisingly, the protocol is strongly linked with counter propagating jams (traffic waves), commonly found in traffic models \cite{Nagel1992,Physicsoftraffic}.

\section{Model and main results \label{sec:Model and main results}}

Let us first introduce the model under study in Sec.~\ref{sec:Model_setup}, followed by summarizing the main results of this work in  Section~\ref{sec:Results}.

\subsection{Model and setup}
\label{sec:Model_setup}
Consider a one-dimensional diffusive many-body systems confined on a ring, whose circumference is set to one. We denote by $\rho(x,t)$, $j(x,t)$ the density and current of particles at position $x\in \left[0,1 \right)$  and time $t\geq 0$.   The density profile and current  satisfy the following hydrodynamics equations \citep{Spohn1992,Bertini2015}: \begin{eqnarray}
\label{eq:diffusion evolution equations}
\partial_ t \rho (x,t)&=& -\partial_x j(x,t)  
\\ \nonumber
  j(x,t) &=& - D(\rho) \partial_x \rho + \sigma(\rho) F(x,t),
\end{eqnarray}
where  $D(\rho)$ and $\sigma(\rho)$ are the diffusion constant and the  conductivity of the particles, determined from microscopic details of the system. $F(x,t)$ is an external force that specifies whether the system is has a NESS or is a SP (precise definitions are given below). This hydrodynamic description holds for a general class of models, for which mathematically rigorous methods have been developed \citep{kipnis2013scaling}.  Note that $D(\rho)$ and $\sigma(\rho)$ are connected through the Einstein relation \citep{Derrida2007} 
\begin{equation}
\label{eq:Einstein relation}
    2D/\sigma  = f''(\rho),
\end{equation}
where $f(\rho)$ is the free energy per unit length. Throughout this paper we set the Boltzmann constant and the temperature to unity. The entropy production rate \citep{Bertini2015,Seifert2005,Busiello2018} is given by  
\begin{equation} 
\label{eq:EP definition}
\Sigma = 2\int dx \,  j(x,t) F(x,t) = 2 \int dx \,  \frac{j^2(x,t)}{\sigma(\rho(x,t))},
\end{equation}
which measures ``how far'' the system is from equilibrium, as demonstrated in Appendix~\ref{sec:app:MFT EP}.

 \subsubsection{NESS}

We consider a class of systems where the force is time-independent ($F(x,t)= F(x)$), and is composed of  both conservative and non-conservative forces:
\begin{equation}
F(x) = E  - \frac{\partial U_{\rm ss}(x)}{\partial x},
\end{equation}
with a constant driving force $E$ and a time-independent potential $U_{\rm ss}(x)$. 
The system relaxes to a unique steady state (NESS), where the steady state density profile and the corresponding current are denoted by $\rho_{ss}(x)$ and $j_{ss}$. Note that the steady state current is spatially independent due to the conservation of particles. 
The entropy production rate in this steady state, $\Sigma_{\rm ss}$, is given in \eqref{eq:EP definition}, where $\lbrace j,\rho \rbrace$ is replaced by $ \lbrace j_{\rm ss},\rho_{\rm ss} \rbrace $.
  In what follows, we name  $\rho_{ss}(x)$, $j_{ss}$, and $\Sigma_{\rm ss}$ the fundamental properties  of the system and investigate if  these properties can be ``imitated'' by another class of systems, SPs.

 \subsubsection{SP}
 
  For a SP on a ring, the force $F(x,t)$ is defined as the gradient of a potential that is periodic in both space (period 1) and time (period $T$). Namely, 
 \begin{equation}
 \label{eq:SP potential formal}
    F(x,t) = - \frac{\partial U_{\rm sp}(x,t)}{\partial x},
 \end{equation}
where $ U_{\rm sp}(x+1,t) =  U_{\rm sp}(x,t) $   and $ U_{\rm sp}(x,t+T) =  U_{\rm sp}(x,t)$ for any $x,t$.  After a few cycles (but not independent of the initial conditions), the density $\rho(x,t)$ and the current $j(x,t)$ converge to  time-periodic states (Floquet states) $ \rho_{\rm sp}(x,t),j_{\rm sp}(x,t)$ with period $T$. Using \eqref{eq:diffusion evolution equations}
and \eqref{eq:Einstein relation}, one can show that the definition of the SP force \eqref{eq:SP potential formal} is equivalent to the integral relation
 \begin{equation}
 \label{eq:generalized detailed balance}
     \int dx \, \frac{j_{\rm sp}}{\sigma(\rho_{\rm sp})} =0 \quad \forall t.
 \end{equation}
In the present article, we name this relation the equilibrium force condition  \footnote{Using the master equation approach,  the  ``detailed balance''  term for  \eqref{eq:generalized detailed balance} is more natural. See \citep{Raz2016}}.

The entropy production rate in a SP is denoted by $\Sigma_{\rm sp}$, defined as (\ref{eq:EP definition}) with replacing $\rho,j$ by $\rho_{\rm sp}$, $j_{\rm sp}$. 
Note that $\rho_{\rm sp}$, $j_{\rm sp}$ and $\Sigma_{\rm sp}$ depend on time. To compare these quantities with the fundamental properties in NESS, we thus introduce time averaging over one cycle in this time-periodic state:
\begin{eqnarray}
\overline{\rho_{\rm sp}}(x) = \frac{1}{T} \int_0 ^{T} dt \, \rho_{\rm sp}, \quad \overline{j_{\rm sp}}(x) & =  \frac{1}{T} \int_0 ^{T} dt \, j_{\rm sp}, \\  \nonumber
 \overline{\Sigma_{\rm sp}}  =  \int_0 ^{T} dt \, \Sigma_{\rm sp}.
\end{eqnarray}

\subsection{ Results \label{sec:Results}}

Here we present the limitations of designing a SP protocol to imitate the fundamental properties of a NESS system.  
The derivation of the results follows in the next section.

 \begin{figure*}
     \centering
     \includegraphics[width=0.3\textwidth]{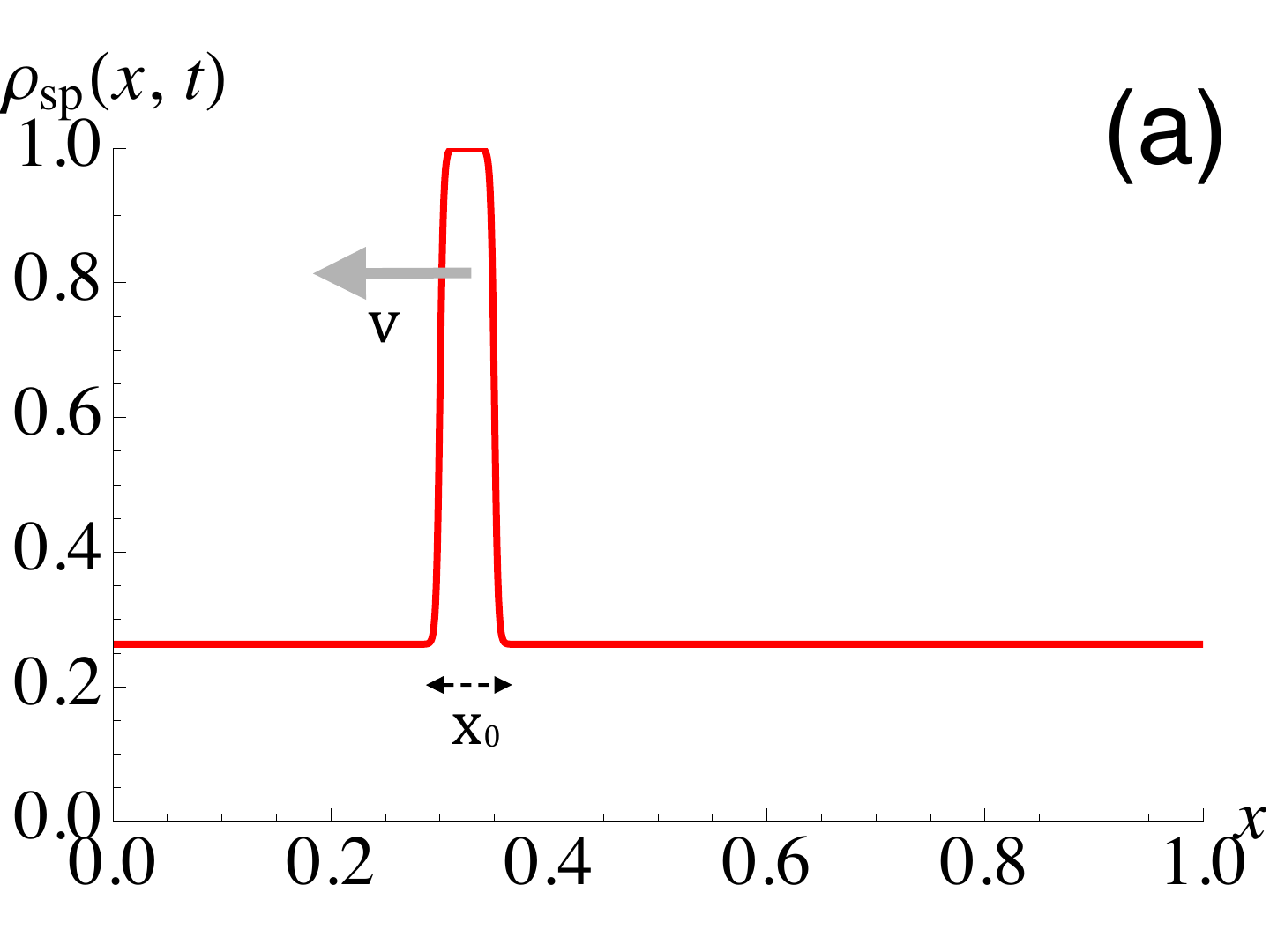} 
     \includegraphics[width=0.3\textwidth]{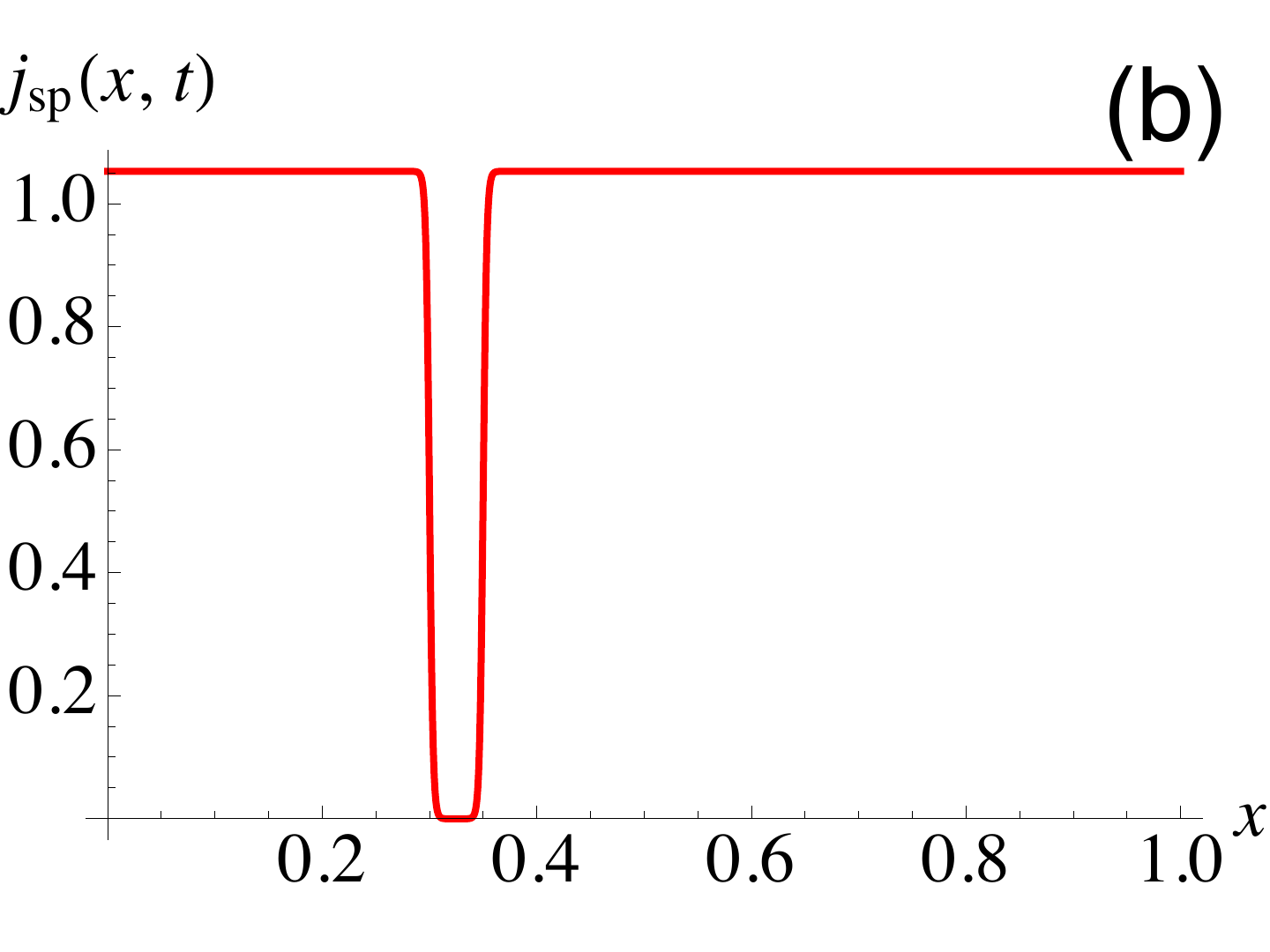} 
     \includegraphics[width=0.3\textwidth]{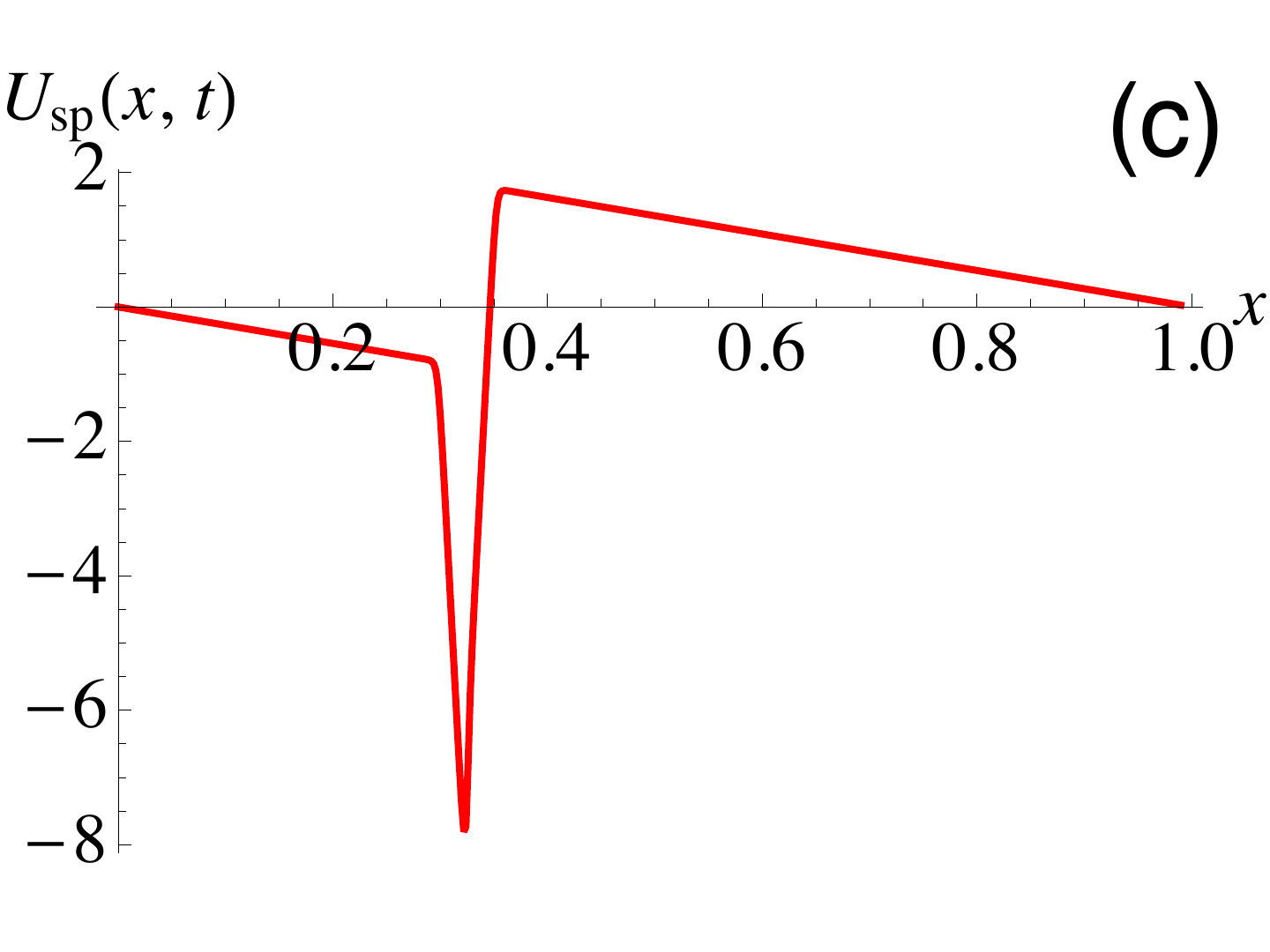} 
\caption{
     An numerical example of the travelling wave SP protocol (\ref{eq:density sp_0}), $\rho_{\rm sp}$ (a), $j_{\rm sp}$ (b), and $U_{\rm sp}$ (c) saturating the entropy production rate bound.  For illustration, we consider the Simple Exclusion Process (SEP) \cite{footnote1}, {\it i.e.}, $\sigma(\rho) = 2 \rho (1-\rho)$ and $D=1$. For the NESS, we set $j_{\rm ss}=1$, $U_{\rm ss}=0$, and $\rho_{\rm ss}=0.3$. We use hyperbolic tangent functions to define the step function $H_{x_0}$ (see Appendix \ref{Section_Numerical} for more details) with setting $x_0=0.05$.  $j_{\rm sp}$ and $U_{\rm sp}$ are determined from $\rho_{\rm sp}$ as detailed in the text. The time-averages, $\overline{\rho_{\rm sp}}$ and $\overline{j_{\rm sp}}$, are exactly equal to $\rho_{\rm ss}$ and $j_{\rm ss}$, whereas the entropy production rate $\overline{\Sigma_{\rm sp}}$ is greater than $\Sigma_{\rm ss}= 2 j_{\rm ss}^2/\sigma(\rho_{\rm ss})$ due to the bound (\ref{eq:EP bound}). However, this bound can be saturated {\it i.e.}, the difference between these entropy production rates  can be arbitrary small by decreasing $x_0$. 
     }
     \label{fig:heaviside}
 \end{figure*}

\subsubsection{Entropy production bound}

First, we generalize the limitations on the entropy production rate found in \citep{Busiello2018}  for a single Brownian particle.  Consider a NESS given by $\rho_{ss}(x),j_{ss}$ 
with a concave conductivity, i.e. $\sigma''(\rho)\leq 0$.  In this case, all the SP protocols with the same $D,\sigma$ that imitate the current and density profile 
of the NESS system, {\it i.e.},
\begin{eqnarray}
\label{eq:imitate rho j }
\overline{ \rho_{\rm sp}}=  \rho_{\rm ss}  & \quad &
\overline{ j_{\rm sp}}=j_{\rm ss},
\end{eqnarray}
result in higher entropy production rate than that of the NESS system, 
 \begin{equation}
 \label{eq:EP bound}
\overline{ \Sigma_{\rm sp}} > \Sigma_{\rm ss}.
\end{equation}
 Note that many  models lead to  concave conductivity, e.g. non-interacting random walkers with $D=1,\sigma=2\rho$ \citep{Busiello2018} and the simple exclusion process (SEP) with $D=1,\sigma=2\rho(1-\rho)$ \cite{Mallick2015}.

\subsubsection{Saturation of the bound}

The saturation of the bound (\ref{eq:EP bound}) is only possible asymptotically as the inequality is strict. For models with strictly concave conductivity, we obtain a necessary condition for this asymptotic saturation. Assume that the protocol depends on control parameter $\lambda$, denoted by $\rho^\lambda _{\rm sp}, j^\lambda _{\rm sp}$, with constraints
\begin{eqnarray}
\label{eq: density current equality}
\overline{ \rho^\lambda _{\rm sp}}=  \rho_{\rm ss},  & \quad &
\overline{ j^\lambda _{\rm sp}}=j_{\rm ss} \quad  \ \ \forall   \lambda.
\end{eqnarray}
If the inequality is saturated in a certain limit ({\it e.g.}, $\overline{\Sigma^\lambda _{\rm sp}} \rightarrow \Sigma_{\rm ss}$  as $\lambda \rightarrow 0$), we then show that the protocol density $\rho^\lambda _{\rm sp}$ needs to converge to the NESS density $\rho_{\rm ss}$ in the same limit, almost everywhere. This necessary condition severely restricts the degrees of freedom we need to explore when designing the saturating protocol.

Let us consider homogeneous NESS driving  -- $U_{\rm ss}(x) =0$, as well as processes with  bounded density, {\it i.e.},  $\rho \in \left[ 0 ,\rho_{\rm max}\right)$ that satisfies $\sigma(\rho_{\rm max})=0$. In this case  we can then design a universal saturating protocol, independent of the exact shape of $D(\rho)$ and $\sigma(\rho)$. The protocol constitutes of a travelling wave of a jammed region that moves to the opposite direction of the particle current
\begin{equation}
\label{eq:density sp_0}
    \rho_{\rm sp}(x,t) = \rho_{\rm ss} + H_\lambda (x-t v),
\end{equation}
where $v=1/T$ and $H_{\lambda}(x)$ is a step function that describes a jammed region (whose precise definitions are given as \eqref{eq:H definition} or \eqref{eq:heaviside}). The width of the jammed region $x_0$ is a control parameter of our SP protocol and the small $x_0$ limit results in saturating the bound. A numerical example of the SP density, current and potential is given in Figure \ref{fig:heaviside}.

Note that a different protocol to saturate the bound (\ref{eq:EP bound}) was proposed in \citep{Busiello2018} for non-interacting particles ($\sigma=2\rho$). However this protocol  cannot be applied to interacting particles with bounded density. See  Appendix.
\ref{sec:app: general models delta protocol} for more details.



\section{Derivation \label{sec:Derivation}}

In this section, we formally derive the results reported in the previous section. In Section \ref{sec:deri_bound}, we prove the entropy production rate bound \eqref{eq:EP bound}, and in Section \ref{sec:deri_saturation} we derive the protocol proposed in \eqref{eq:density sp_0} to saturate the bound. 
In what follows, the shorthand notation $\sigma_{X} = \sigma( \rho_{\rm X})$  for $X = \lbrace ss,sp \rbrace $ is used.

\subsection{Derivation of the bound}
\label{sec:deri_bound}

Consider a NESS and a SP systems with the same diffusive dynamics $D,\sigma$. Suppose that the SP is designed to follow \eqref{eq:imitate rho j }. Let us define

\begin{eqnarray}
\label{eq:R_macro}
R = \frac{2}{T}\int dx dt \, \sigma_{sp} \left(  \frac{j_{sp}}{\sigma_{sp}} - \frac{j_{ss}}{\sigma_{ss}}  \right)^2. 
\end{eqnarray}
When the conductivity $\sigma$ is concave, expanding \eqref{eq:R_macro} and using \eqref{eq:imitate rho j } reveals that  $R\leq \Delta \Sigma \equiv \overline{\Sigma_{sp}}  - \Sigma_{ss}$.  We thus conclude $\Delta \Sigma \geq 0$. Note that an equality  $\Delta \Sigma = 0 $ can only be achieved for   $\frac{j_{sp}}{\sigma_{sp}}= \frac{j_{ss}}{\sigma_{ss}}$ $\forall x,t$. However,  the equilibrium force condition \eqref{eq:generalized detailed balance}  implies that this cannot be possible for a NESS system as 
\begin{eqnarray}
     \int dx \, \frac{j_{sp}}{\sigma_{sp}} = 0, & \quad {\rm whereas} \quad & 
       \int dx \, \frac{j_{ss}}{\sigma_{ss}} \neq 0. 
\end{eqnarray}
Thus we have verified  \eqref{eq:EP bound}  for $\sigma $ concave models. 
This result implies that any SP protocol designed to have the same time-averaged density and current will always generate a higher (time-averaged) entropy production rate   than the NESS process. Note that for processes with non-concave $\sigma$, there is no such entropy production rate bound \eqref{eq:EP bound}. See the appendix \ref{sec:app: general models delta protocol} for an example.


\subsection{Saturating protocol }
\label{sec:deri_saturation}

We consider protocols $\rho_{\rm sp}^{\lambda}$ that depend on the set of parameters $\lambda=(\lambda_1,\lambda_2,...)$ where we assume 
\begin{eqnarray}
\overline{\rho^\lambda _{\rm sp}}=\rho _{\rm ss},  
\, \overline{j^\lambda _{\rm sp}}= j _{\rm ss}& \quad \quad & \forall \lambda_i>0.
\end{eqnarray}
Among these protocols, we search for one that saturates the bound as all the $\lambda_i$'s tend to vanish, {\it i.e.}
\begin{eqnarray}
\lim_{\lambda_i \rightarrow 0, \forall i}  \overline{\Sigma^\lambda _{\rm sp}} = \Sigma _{\rm ss}.
\end{eqnarray}
From here on out, we focus only on concave $\sigma$ models with a bounded density.


\subsubsection{No go protocols \label{sec:No go protocols}  }

Before we present how to design a SP protocol that imitates the NESS system, it is important to learn which protocols {\it cannot}. For this purpose, let us study in detail the source of the entropy production bound \eqref{eq:EP bound}. 
For strictly  concave  $\sigma$, $d(x) \equiv \sigma_{ss} - \overline{\sigma_{sp}}$  is greater than zero except for the case where $\rho_{sp}(x,t) = \rho_{ss}(x)$ for any $t$. Then, since 
\begin{equation}
0 \leq R = \Delta \Sigma - 2 \int dx \frac{j^2 _{ss}}{\sigma^2 _{ss}} d(x)     
\end{equation}
by definition, 
we find 
\begin{equation}
\label{eq:bounding the EP}
\Delta \Sigma \geq G \int dx \, d(x) \geq 0,      
\end{equation}
for $j_{ss}\neq 0$, where $G = \min_x \frac{j^2 _{ss}}{\sigma^2 _{ss}}  >0$. The entropy production rate bound can be saturated only if
$\int dx \, d (x)   \rightarrow 0$ as $\lambda \rightarrow 0$. This means that our protocols need to satisfy $d (x) \rightarrow 0$ (and hence $\rho^\lambda _{sp}(x,t)\rightarrow \rho_{ss}(x)$) as $\lambda \rightarrow 0$ almost everywhere on the ring.  Demanding $d (x) \rightarrow 0$ everywhere leads to breaking the equilibrium force condition \eqref{eq:generalized detailed balance}, as shown in Appendix \ref{sec:app:close but not}. The only possibility for our protocols is thus to force out  $\rho^\lambda _{sp} \rightarrow \rho_{ss}$ as $\lambda \rightarrow 0$, but only {\it almost} everywhere.

Let us denote the spatial region  where $\rho^\lambda _{sp} \rightarrow \rho_{ss}$ by $\Omega^\lambda \subset [0,1]$ and its complement on the ring by $\Omega^{\lambda,\perp} $. 
To satisfy the equilibrium force condition \eqref{eq:generalized detailed balance}, we need to balance the two parts of the integral, namely,  $\int_{\Omega^\lambda} j^\lambda _{sp}/\sigma^\lambda _{sp}$ and $-\int_{\Omega^{\lambda,\perp} } j^\lambda _{sp}/\sigma^\lambda _{sp}$. 
At the small $\lambda$ limit, $\Omega^\lambda \rightarrow [0,1]$ and $\Omega^{\lambda,\perp} $ becomes vanishingly small. Therefore,  $|j^\lambda _{sp}/\sigma^\lambda _{sp}|$ must diverge in $\Omega^{\lambda,\perp} $ for  $\lambda \rightarrow 0$ for the two integrals to be balanced.  
A possible scenario for such a divergence is for  $\sigma^\lambda _{sp} \rightarrow 0$ 
in $\Omega^{\lambda,\perp}$. In concave models with a bounded density $\rho_{\rm max}$, the conductivity vanishes as $\rho\rightarrow 0,\rho_{\rm max}$ \footnote{Note that we may also try to force the current to diverge at the limit.}. Here, we will consider having $\rho \rightarrow \rho_{\rm max}$  in the small domain $\Omega^{\lambda,\perp}$ to saturate the bound. This implies that the protocol creates a jammed region that  propagates to the opposite direction of the current (see Figure \ref{fig:heaviside}). This can be inferred by an old argument by Stokes  using the continuity equation \eqref{eq:diffusion evolution equations}. See Sec.3.2.1. of  \citep{Physicsoftraffic}.


\subsubsection{Saturation of the bound}
\label{sec:derivation:saturation}

So far we have discussed necessary conditions for a protocol to saturate the bound \eqref{eq:EP bound}. One possibility for the protocol is to generate a small jammed region counter propagating to the direction of  the particle current. 
Here we detail how to precisely design this protocol and show the saturation of the bound in a special case. Recall we consider systems with bounded density $\rho_{\rm max}$, {\it i.e.}, $\sigma(\rho_{\rm max})=0$, (below we set $\rho_{\rm max}=1$ without loss of generality). We consider the NESS system that is driven by a uniform  force $F(x)=E$, leading to a homogeneous steady state  density profile $\rho_{\rm ss}$ and current $j_{\rm ss}$. To imitate that, we consider the travelling wave protocol \eqref{eq:density sp_0}, where $H(x)$ is a periodic function with period 1, defined   as 
\begin{equation}
\label{eq:H definition}
     H_{x_0,\epsilon}(x)=\left\{
                \begin{array}{ll}
                  1-\rho_{ss}-\epsilon  & 0\leq x\leq x_0 \\
                  C  &  x_0 \leq  x <1,
                \end{array}
              \right.
\end{equation}
where $x_0$ ($0<x_0<1$) is the width of the step function introduced in the previous section and $\epsilon>0$ is an additional parameter to avoid a complete jam of particles at the higher density region. This latter parameter is technically required to avoid a fictitious divergence in the equilibrium force condition \eqref{eq:generalized detailed balance} and we always assume $\epsilon \ll x_0$ below.  The constant $C$ is determined so as to satisfy the matching condition of the densities $\overline{\rho_{\rm sp}} = \rho_{\rm ss}$: $C=-\frac{x_0}{1-x_0} (1-\rho_{ss}-\epsilon)$. For convenience of notations, let us also drop the dependence of $j_{\rm sp},\rho_{\rm sp}$ in $\lambda = (x_0,\epsilon)$. 
As for the current,  the form \eqref{eq:density sp_0}, together with the continuity equation \eqref{eq:diffusion evolution equations} and the equilibrium force condition \eqref{eq:generalized detailed balance} determine the SP current 
\begin{eqnarray}
\label{eq: jSP def alpha def}
j_{sp}(x,t) &=& v \left(  \rho_{sp} - \frac{\alpha_0}{\alpha_1}  \right) \end{eqnarray}
with time-independent constants
\begin{eqnarray}
\alpha^{-1} _k &=&  \int dx \, \frac{ \rho_{sp}^k }{\sigma_{sp}}.
\end{eqnarray}
The time-averaged SP current can then match $j_{\rm ss}$ by setting the velocity (equivalently the cycle time $T=1/v$) of the travelling wave to 
\begin{equation}
v = \frac{\overline{j_{\rm sp}}}{\rho_{\rm ss} - \alpha_0/\alpha_1}, 
\end{equation}
which can be derived from (\ref{eq: jSP def alpha def}). In these settings, the saturation of the bound can be proven as follows: the entropy production rate in the SP  is 
\begin{eqnarray}
\label{eq: flat profile current alpha}
\overline{\Sigma_{\rm sp}} = \Sigma_{\rm sp}   &=&  2 v^2 (\frac{1}{\alpha_2 } - \frac{\alpha_0}{\alpha^2 _1}).
\end{eqnarray}
On the other hand, the entropy production rate in the NESS is  $\Sigma_{ss} = 2  j^2 _{ss}/ \sigma_{ss}$. To compare these two entropy production rates we use 
\begin{equation}
    \sigma_{sp} \approx 
    \left\{
                \begin{array}{ll}
                  \sigma(1-\epsilon)  & 0\leq x -vt\leq x_0 \\
                  \sigma_{ss}  &  x_0 \leq  x-vt<1, 
                \end{array}
              \right.
\end{equation}
up to $O(\epsilon,x_0)$ corrections. We further expand $\sigma(1-\epsilon)\approx -\sigma'(1)\epsilon + O(\epsilon^2)$ using $\sigma(1)=0$. These expansions allow us to evaluate $\overline{j_{\rm sp}}$ and $\overline{\Sigma_{\rm sp}}$ analytically, 
\begin{eqnarray}
\overline{j_{\rm sp}} &=& - v \left (1- \rho_{\rm ss}\right ) + C_0 \frac{\epsilon}{x_0}  + O(x_0,\epsilon)
\\ \nonumber
\overline{\Sigma_{\rm sp}} &=& v ^2 \left (1- \rho_{\rm ss}\right )^2  \frac{2}{\sigma_{\rm ss}} + C_1 \frac{\epsilon}{x_0}  + O(x_0,\epsilon)
\end{eqnarray}
with constants $C_0$ and $C_1$. This implies that taking first $\epsilon \rightarrow 0 $ and then $x_0 \rightarrow 0$ saturates the bound: $\overline{\Sigma_{\rm sp}} \rightarrow 2 \overline{j_{sp}}^2 /\sigma_{ss} = \Sigma_{\rm ss}$.




\section{Discussion \label{sec:Discusion}}

A bound for the entropy production rate exists for SP protocols \eqref{eq:EP bound} when trying to mimic the density profile and current of a NESS system \eqref{eq:imitate rho j }, provided that the conductivity of the model has a concave form. For homogeneous NESS driving, this bound can be saturated using a special protocol in which a travelling wave propagates against the direction of the particle current (Figure~\ref{fig:heaviside}). These are non-trivial extensions of the known results obtained in \citep{Busiello2018} for a single Brownian particle system to many-body systems. We stress that (i) the concavity for the bound has not been discussed in \citep{Busiello2018} as it does not play any important role for their single particle case and (ii) their proposed saturating protocol approximating the Dirac $\delta$ function \citep{Busiello2018} cannot be applied to arbitrary many-body systems. In Appendix \ref{sec:app: general models delta protocol}, we show how different NESS settings for a convex concavity model can lead also to  $\Sigma_{\rm ss} \geq \overline{\Sigma_{\rm sp}}$ and how the $\delta$ function protocol is unfeasible for models with bounded density, e.g. exclusion processes.

The bound for entropy production rates is specific to the concave conductivity models on a continuous space. In the discrete space  master equation approach,  the corresponding bound does not exist \citep{Raz2016}. In this case, there exists a protocol that always imitates the SP fundamental properties  to the NESS ones. We point out two major differences between the discrete and the continuous cases. First, in the discrete case, the NESS master equation is uniquely determined by the currents, steady state probabilities and entropy production rates between each two sites. However, for the diffusive continuous case, the probability and entropy production rate are dominated by the steady state current and steady state density profile. Corrections due to fluctuations in the system are exponentially suppressed. Therefore, we have in principle less degrees of freedom. Determining the current and probability determines the entropy production uniquely. Secondly, we restricted the SP system  imitating the NESS system with similar $D,\sigma$. Namely, in the continuous description, we kept in tact the inter-particle interactions. In the discrete case, complete freedom was allowed for changing the transition rates in the SP case, provided that they satisfy the equilibrium force condition. Allowing different $D,\sigma$ in the SP  completely nullifies the existence of the entropy production rate bound.

Designing a protocol for inhomogeneously driven NESS systems is left as a future challenge. The arguments of Sec.~\ref{sec:No go protocols} suggest that if such a protocol exists, it is likely to have a similar counter propagating jam-like form. 
It is possible that for higher dimensional systems as well as systems with different boundary conditions similar ideas allow to design a saturating protocol. It will be interesting also to generalize the above results in the case of non-diffusive systems.

The unexpected role that propagating traffic jams play in  minimizing  the entropy production rate in a SP is worth further exploration. Traffic waves are ubiquitous in the study of highway traffic \cite{kerner2012physics}.   It is interesting to understand whether one can explain the formation of traffic waves using a minimum entropy production rate principle \citep{Nagel1992,AppertRolland2011}. 
The usual definition of entropy production rate relies on the non-zero probability for any  time-reversed trajectory (see Appendix~\ref{sec:app:MFT EP}). However, many traffic models do not respect this assumption as cars tend not to move backwards on highways. With this regard, a technique that replaces second derivatives in time-evolution equations by convective derivatives \cite{Aw_Rescle_2000} could be useful. Furthermore, we could rely on a recent publication \citep{Busiello2019} that shows how the definition of entropy production rates can be extended to the cases where time-reversed trajectories may be prohibited.

Finally, we note that molecular machines face many challenges, where minimizing their entropy production rate may not be a primary concern. For example, in small systems, the typical and mean behaviour is not always the same. Then, it becomes relevant to focus on the dominant fluctuations and not only the mean as we have done here.  It would be interesting to compare  NESS and SP where other constraints arise and see whether similar universal bounds are at play. Note that we  have used the hydrodynamic approach to uncover the entropy production rate bound. The hydrodynamic approach, assuming large systems, seems at first glance ill-equipped to handle systems with large fluctuations. However, as we have seen in this work, it is common that the clean approach of the hydrodynamic equations allow to notice such universal bounds, where microscopic approaches did not.


\begin{acknowledgments}
The authors acknowledge fruitful discussions with C\'ecile Appert-Rolland, Bernard Derrida, Henk Hilhorst,
 Patrick Pietzonka and Oren Raz.   
\end{acknowledgments}

\appendix


\section{The entropy production rate expression \label{sec:app:MFT EP}}

In this Appendix, we derive the standard expression for the entropy production rate in diffusive systems \citep{Bertini2006}.

 The time-evolution equations \eqref{eq:diffusion evolution equations} of the diffusive density profile are derived in a hydrodynamic (coarse graining) limit from microscopic models. These equations are deterministic and describe the most likely evolution. Small fluctuations from the most likely evolution exist in macroscopic regimes and they can be described by {\it fluctuating hydrodynamics} \citep{Bertini2015,Shpielberg2016}, given in terms of the Langevin-like equation 
\begin{eqnarray}
\partial_t \rho(x,t) &=& - \partial_x j (x,t)
\\ \nonumber 
j(x,t) &=& J(\rho) +  \sqrt{\sigma(\rho)/L} \, \xi(x,t)
\\ \nonumber 
J(\rho) &=& -D(\rho) \partial_x \rho + \sigma (\rho) F, 
\end{eqnarray}
where $L$ denotes the (large) system size defined in the microscopic model and $\xi(x,t)$ is a Gaussian white noise with zero mean. One can use a Martin-Siggia-Rose procedure \footnote{The small noise allows to disregard Ito vs. Stratonovich questions \cite{Aron_2016,Itami2017}.} to evaluate the probability of observing a trajectory $\lbrace \rho(x,t),j(x,t) \rbrace$ at the time window $t\in [0,T]$ and for $x\in [0,1]$. 
The probability is given by 
\begin{equation}
    P(\lbrace\rho,j\rbrace) \sim \exp{ -L \int dx dt  \, 
    \frac{(j- J(\rho) )^2 }{2\sigma }  } ,
\end{equation}
where sub-exponential contributions are discarded \footnote{In principle the probability expression is usually accompanied by the quasi-potential accounting for the the probability to find the system in a given initial state \citep{Bertini2015}. The entropy production rate will not be affected by the subtraction of the quasi-potential when the periodic boundary conditions are considered.   }.

The entropy production  between $t=0$ and $t=T$ for a given trajectory $\lbrace \rho(x,t),j(x,t) \rbrace$, denoted by
$\mathbb{S}(\lbrace \rho,j \rbrace )$,
compares the probability of the trajectory and its time reversed trajectory. Namely, 
\begin{eqnarray}
\mathbb{S}(\lbrace \rho,j \rbrace )
=  \log \frac{P(\lbrace \rho,j \rbrace )}{P(\lbrace \theta \rho, \theta j \rbrace )},
\end{eqnarray}
 where $\theta$ is the time-reversal operator. It gives according to the macroscopic theory 
\begin{eqnarray}
\mathbb{S} (\lbrace j,\rho \rbrace ) = -L \int dx  dt\,  \frac{2 j}{\sigma} (D\partial_x \rho - \sigma F ). 
\end{eqnarray}
The  entropy production rate is given by averaging over all the trajectories. Namely 
\begin{equation}
\Sigma  = \frac{1}{T L } \int \mathcal{D} j\mathcal{D} \rho \,  P(\lbrace \rho,j \rbrace) \mathbb{S}(\rho,j) . 
\end{equation}
The probability is dominated by a single trajectory that satisfies $j = J(\rho)$ together with the continuity equation. This implies  
\begin{eqnarray}
\Sigma =  2 \int dx \, \frac{J^2}{\sigma} 
\end{eqnarray}
that corresponds to the expression \eqref{eq:EP definition} in the main text.

\section{Hyperbolic tangent function for rounded step function}
\label{Section_Numerical}

In this Appendix, we introduce a rounded step function used in Figure~\ref{fig:heaviside} to express a travelling wave $H_{\lambda}$ of (\ref{eq:density sp_0}). Different from the other travelling function defined as (\ref{eq:H definition}), this step function is continuously differentiable, so that we can take its derivative to obtain the corresponding SP potential shown in Figure~\ref{fig:heaviside}.

\begin{figure}
     \centering
     \includegraphics[width=0.4\textwidth]{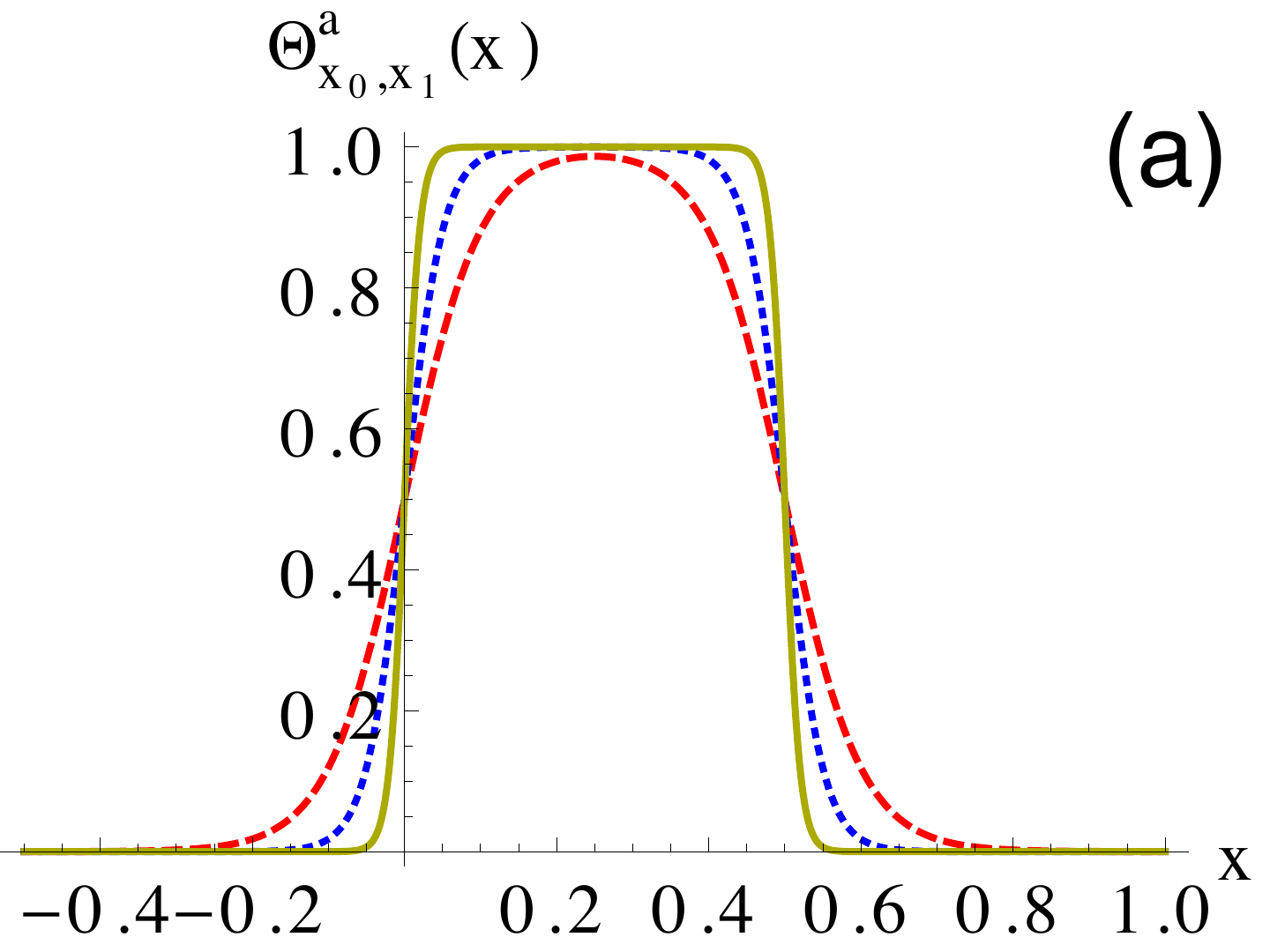} 
     \includegraphics[width=0.4\textwidth]{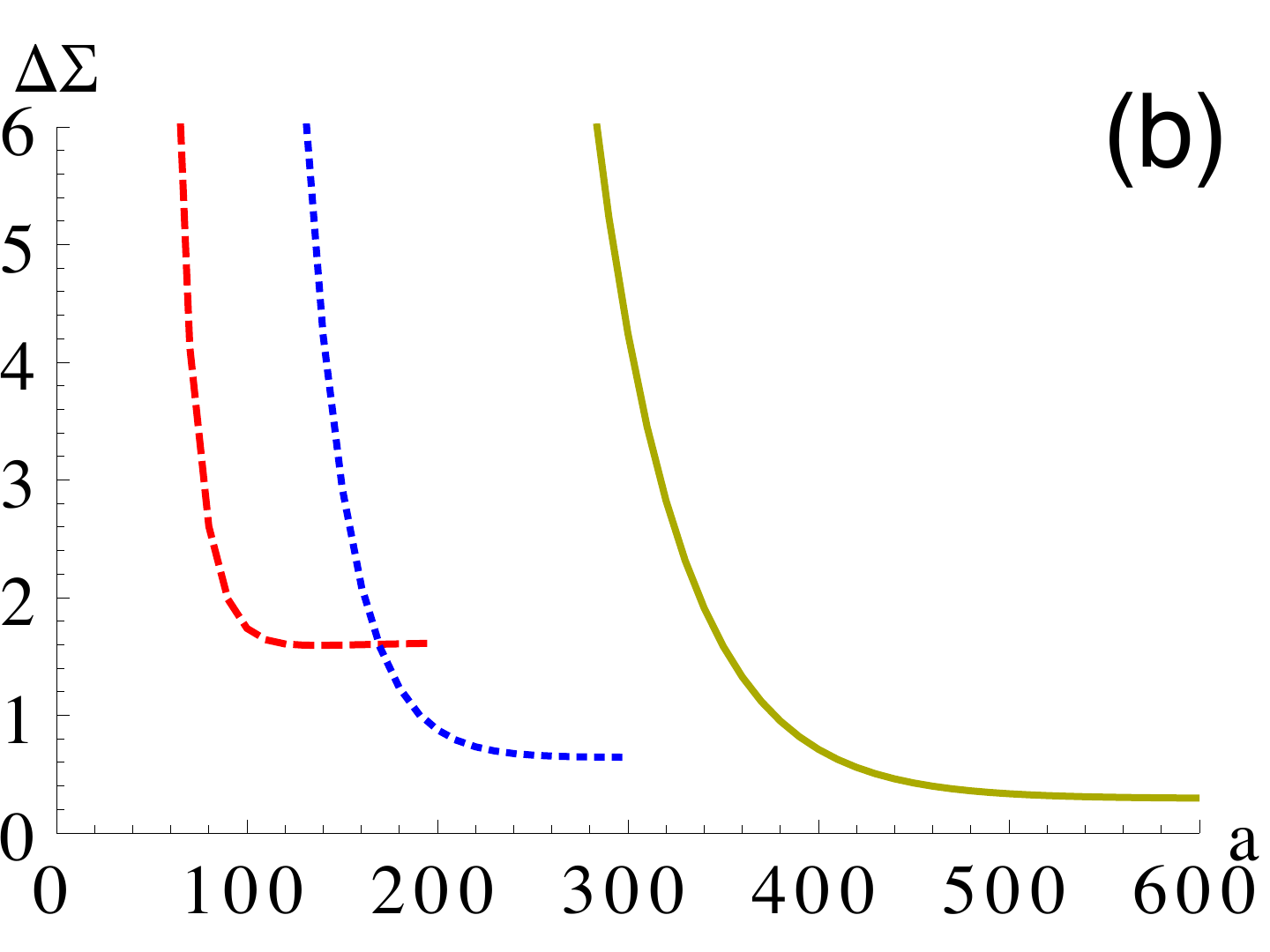}
     \caption{({\bf a}) The rounded step function  $\Theta^a_{x_0,x_1}(x)$ defined in (\ref{eq:heaviside}) for several values of $a$ (Dashed red: $a=10$, Dotted blue: $a=20$, Solid yellow: $a=50$). Here we set $x_0=0$ and $x_1=1$. The function converges to a step function as $a$ increases. 
     ({\bf b}) The difference of the entropy production rates $\Delta \Sigma \equiv  \Sigma_{\rm sp} - \Sigma_{\rm ss}$ as a function of the rounding parameter $a$ in SEP. Different colors represent different values of $x_0$ (Dashed red: $x_0=0.1$, Dotted blue: $x_0=0.05$, Solid yellow: $x_0=0.025$). We set $\rho_{\rm ss}=0.3$ and $j_{\rm ss}=1$. The figure demonstrates the saturation of the bound as both $x_0$ and $1/a$ decrease.
     }
     \label{fig:appendix:saturation}
 \end{figure}

First, we introduce the following combinations of hyperbolic tangent functions with parameters $x_0,x_1,a$, 
\begin{equation}
\Theta_{x_0,x_1}^a(x) = -1 + \frac{1}{1 + e^{-2 a (x-x_0) }} + \frac{1}{1 + e^{2 a (x - x_1)}},
\end{equation}
which converge to a step function from $x_0$ to $x_1$ in the large $a$ limit (Figure~\ref{fig:appendix:saturation}). 
We then define the rounded periodic step function 
\begin{equation}
\label{eq:heaviside}
H^{\rm hyp}_{x_0,a}(x) = (1-\rho_{\rm ss}) \Theta_{0,x_0}(x) - \Theta_{x_0,1}(x) \frac{ x_0(1 - \rho_{\rm ss})}{1 - x_0}.
\end{equation}
As this function is continuously differentiable, the SP potential $U_{\rm sp}$ can be calculated from the hydrodynamic equation (\ref{eq:diffusion evolution equations}) \begin{equation}
    \partial_x U_{\rm sp}(x,t)= - \left[ j_{\rm sp}(x,t) + D_{\rm sp}  \partial_x \rho_{\rm sp}(x,t) \right ]\frac{1}{\sigma_{\rm sp} },
\end{equation}
where $\rho_{\rm sp}$ and $j_{\rm sp}$ are given in (\ref{eq:density sp_0}) and (\ref{eq: jSP def alpha def}). 
These three functions, $\rho_{\rm sp}$, $j_{\rm sp}$ and $U_{\rm sp}$ are evaluated for the  homogeneously driven SEP with $a=300$ and plotted in Figure~\ref{fig:heaviside}.
  In the same model, we also compare the SP entropy production rate $\Sigma_{\rm sp}$, (\ref{eq: flat profile current alpha}), with the NESS entropy production rate, $\Sigma_{\rm ss}=2 j_{\rm ss}^2 /\sigma_{\rm ss}$. In Figure~\ref{fig:appendix:saturation}, we plot $  \Delta \Sigma \equiv  \Sigma_{\rm sp} - \Sigma_{\rm ss}$ as a function of $a$ for several values of $x_0$. The bound is saturated as both $x_0$ and $1/a$ decrease. Note that the parameter $1/a$ in this protocol and the parameter $\epsilon$ in the protocol corresponding to (\ref{eq:H definition}) play a similar role. Both parameters govern the shapes of the travelling waves, which  converge to the same step function by decreasing these parameters to zero. In both cases, we first take these limits and then decrease the width of the travelling wave $x_0$ in order to saturate the bound.

\section{No equilibrium force  for $d(x) \to 0$  \label{sec:app:close but not}}

Here we provide a  proof for Section~\ref{sec:No go protocols}. We show that demanding $d(x) \to 0$ for {\it all} $x$ necessarily violates the equilibrium force condition (\ref{eq:generalized detailed balance}).

Let us assume that $\rho^{\lambda} _{sp}(x,t) \rightarrow  \rho _{ss}(x)$ for any $x,t$ as $\lambda \rightarrow 0$. This implies that we can set 
\begin{equation}
    \rho^{\lambda} _{sp}(x,t) = \rho_{ss}(x) + O(\lambda),
\end{equation}
where we use freely the notation $O(\lambda)$ to indicate a function that converges to zero as $\lambda \rightarrow 0$. We further  assume $\rho^\lambda _{\rm sp}-\rho_{\rm ss}$ is smooth in the limit $\lambda\rightarrow 0$ for a reasonable SP generated density profile. The continuity equation suggests that
\begin{equation}
    j^\lambda _{sp}(x,t) = C(t) + O(\lambda), 
\end{equation}
where $C(t)$ is a time-dependent function. A perturbative analysis in $\lambda$ implies that 
\begin{equation}
    \frac{j^\lambda _{sp}}{\sigma^\lambda _{sp}} = \frac{C(t)}{\sigma_{ss}} + O(\lambda).
\end{equation}
The equilibrium force condition \eqref{eq:generalized detailed balance} suggests that $C(t)$ must identically vanish. In this case, we find  that
 $j_{\rm ss}= \overline{j^\lambda _{sp}} =  O(\lambda) $. This makes $j_{ss}$ either $\lambda$ dependent or zero in contradiction to \eqref{eq: density current equality}. Therefore, we cannot have \eqref{eq: density current equality} and $\rho_{sp}$ becoming arbitrarily close to $\rho_{ss}$ everywhere in $x$ at the limit without violating the equilibrium force condition.


\section{The  $\delta$ wave protocol applied to general diffusive systems  \label{sec:app: general models delta protocol}}

Designing a saturating protocol for non-interacting particles was already achieved in \citep{Busiello2018}  using a Fokker-Planck approach for a single particle. Here, we recap the relevant points of this approach in the hydrodynamic framework. Then we analyze the entropy production rate differences between the NESS and SP systems using the $\delta$ wave protocol for two relevant models -- non-interacting diffusing particles and the KMP diffusive heat transport model \citep{Kipnis1982}.  We demonstrate that the $\delta$ wave protocol generally does not lead to $\overline{\Sigma_{\rm sp}}\rightarrow \Sigma_{\rm ss}$.

\subsection{The $\delta$ wave protocol }

Restricting the discussion to a ring system allows to define the globally conserved mass $Q$. Namely, 
\begin{equation}
Q = \int dx \, \rho_{ss}(x) = \int dx \, \rho_{sp}(x,t)      \quad \forall t.
\end{equation}
According to \citep{Busiello2018}, the protocol dictates a travelling wave solution for   the density profile of the SP, $\rho_{sp}(x,t) =  f(x- \hat{x} (t))$. The function $f(x)$  is a  periodic function with period $1$, specifying the shape of the travelling wave and $\hat{x}(t) $ dictates the propagation velocity. One can show that choosing 
\begin{equation}
\frac{1}{\dot{\hat{x}} (x)} = T \sum_n \mathrm{e}^{2\pi i n x } \frac{\rho_{ss,n}}{f_n}, 
\end{equation}
where $\rho_{ss,n},f_n$ are the $n$-th components of the Fourier decomposition of  $\rho_{ss}(x),f(x)$, allows to satisfy $\overline{\rho_{sp}(x,t)} = \rho_{ss}(x)$. 
The continuity equation implies that 
\begin{equation}
j_{sp}(x,t) = j_{sp}(x,t) + \dot{\hat{x}} (t)  \left[ f(x-\hat{x})-f(-\hat{x}) \right].    
\end{equation}
 Together with the equilibrium force condition \eqref{eq:generalized detailed balance}, we find 
\begin{eqnarray}
j_{sp}(x,t) &=& \dot{\hat{x}}(t)  \left[ 
f(x-\hat{x})-\alpha_0/\alpha_1 \right]
\\ \nonumber 
\overline{j_{\rm sp}} &=&
\frac{1}{T}( Q- \frac{\alpha_0}{\alpha_1}   ) ,
\end{eqnarray}
where 
\begin{equation}
\label{eq:alpha k def}
(\alpha_k)^{-1} \equiv  \int dx \, \frac{f^k (x)}{\sigma(f(x))}. 
\end{equation}
This allows to evaluate the entropy production rate as $\Sigma_{sp}(t) = 2 \dot{\hat{x}}^2  (\frac{1}{\alpha_2}-\frac{\alpha_0}{\alpha^2 _1})$.

We then choose to converge  $f(x)\rightarrow Q \delta (x)$ by taking $f(x) = Q \frac{1}{\sqrt{4\pi \epsilon}} e^{- x^2 / 4\epsilon} $ as $\epsilon \rightarrow 0^+$. A qualitative point of view in terms of the potential reveals a deep well, capturing all the particles. The well moves  with a certain velocity defined by $\dot{\hat{x}}_0$.  
Then, the price to pay for reducing the entropy production rate are sharp gradients in the potential. 
Taking the $\delta$ limit leads to 
\begin{eqnarray}
j_{ss} &=& \frac{Q-\alpha_0/\alpha_1}{T}
\\ \nonumber
\Sigma_{ss} &=& 2 j^2 _{ss}\int dx \, \frac{1}{\sigma_{ss}}. 
\end{eqnarray}
Time averaging over a cycle reveals that 
\begin{equation}
    \overline{\Sigma_{sp}} = \frac{2Q}{T^2} (\frac{1}{\alpha_2} - \frac{\alpha_0}{\alpha^2 _1}) \int dx \, \frac{1}{\rho_{ss}(x)}.
\end{equation}
The $\alpha_k$ values depend on the process at question (specifically at $\sigma$). An important remark is that the $\delta$ wave protocol makes no sense for lattice gas models as they have a bounded density. These processes have a  non-convex $\sigma$. Therefore, while a bound on the mean entropy production rate of the stochastic pump system exists only for concave processes, it typically cannot be saturated using the $\delta$ wave protocol. The non-interacting case is an exception. Note that the non-interacting case, with $\sigma = 2\rho$ is trivially concave, but not strictly concave. Therefore, the argument of  Sec.~\ref{sec:No go protocols} need not apply. 
Indeed, in this case, 
$\alpha_0 \to 0, \alpha_1 \to 2, \alpha_2 \to 2/Q  $ in the $\epsilon \rightarrow 0^+$ limit. This implies that the $\delta$ wave protocol optimizes the entropy production rate as $j_{ss} = Q/T$ and $\overline{\Sigma_{sp}}\rightarrow \Sigma_{ss}$.


\subsection{The KMP process \label{sec:app:KMP}  }

For processes with non-concave $\sigma$, there is no entropy production rate bound. Therefore, one should not expect that the $\delta$ wave protocol generally leads  to $\overline{\Sigma_{sp}}\rightarrow \Sigma_{ss}$. To show that, we consider a (macroscopically) simple dynamics with $D=1,\sigma=2\rho^2$ corresponding  to the KMP process. The rest of the appendix is devoted to showing that applying the $\delta$ wave protocol to the KMP  leads to   $\Delta \Sigma \leq 0 $.

\begin{figure}
\begin{center}
 \includegraphics[width=0.43\textwidth]{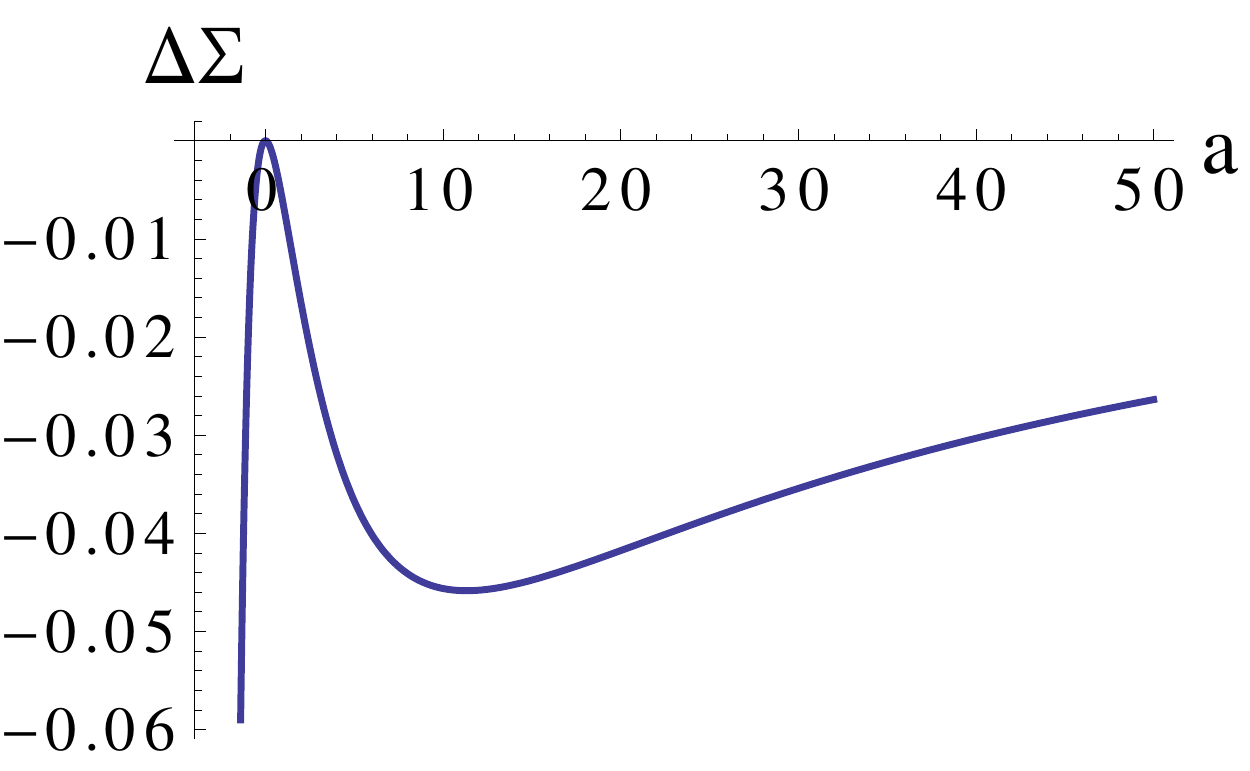} 
\caption{\label{fig:kmp hump profile } 
 The difference of the entropy production rate between the NESS and SP systems, $\Delta \Sigma  = \overline{\Sigma_{\rm sp}} - \Sigma_{\rm ss}$, as a function of $a$. The NESS density profile is $\rho_{\rm ss} (x)= 1+a x (1-x)$ and the NESS current $j_{\rm ss} = 1 $. 
 }
\end{center}
\end{figure}

Using the exponential representation of the $\delta$ function, the KMP dynamics lead to   
$\frac{\alpha_0}{\alpha_1} \to 0, \frac{\alpha_0}{\alpha^2 _1} \to 0, \alpha_2 \to 2  $ in the limit. This implies that 
$\overline{\Sigma_{sp}} = \frac{Q}{T^2} \int dx \, \frac{1}{ \rho_{ss}(x)}$, whereas $\Sigma_{ss} = \frac{Q^2}{T^2} \int dx \, \frac{1}{\rho^2 _{ss} (x)}$. Generally these two do not match. For a periodic system with a constant driving field $E$, we find that the entropy production rate indeed matches as $\rho_{ss} (x) = Q$. 
As an illustrative example, let us consider the density profile $\rho_{ss}(x) = 1 + a x (1-x)$ with the current $j_{ss}= 1$. In this case, one can calculate the  entropy production rate difference $\Delta \Sigma $ 
as a function of $a$ explicitly (See Figure~\ref{fig:kmp hump profile }). We find that   $\Delta \Sigma \leq 0 $ in this case, contrary to the concave case. Moreover, the $\Delta \Sigma \rightarrow 0$ only as $a\rightarrow 0,\infty$ where the system has a constant density profile (which is general for the KMP and probably many other processes) or at the other extreme, where the density gradient is large.

\bibliographystyle{apsrev4-1}   
\bibliography{main.bib} 

\begin{thebibliography}{49}
\bibitem [{foo()}]{footnote1}%
  \BibitemOpen
  \href@noop {} {\bibinfo  {journal} {A SEP is defined as a lattice model.
  Particlle are allowed to jump to empty nearest neighbors. Macroscopically,
  this exclusion rule leads to $D=1,\sigma = 2\rho (1-\rho)$, conforming to a
  strictly concave conductivity, where the density is bounded: $\rho \in
  \left[0,1\right]$}\ }\BibitemShut {NoStop}%
\end{thebibliography}

\end{document}